\documentclass[onecolumn]{IEEEtran}
\usepackage{mathpazo}
\usepackage{times}
\usepackage{amsmath}
\usepackage{amsfonts}
\usepackage{latexsym}
\usepackage{amssymb}
\usepackage{cite}
\usepackage[ruled,lined]{algorithm2e}
\usepackage{upref}
\usepackage{theorem}
\usepackage{graphicx}

\ifCLASSOPTIONcompsoc
\usepackage[caption=false,font=normalsize,labelfont=sf,textfont=sf]{subfig}
\else
\usepackage[caption=false,font=footnotesize]{subfig}
\fi

\usepackage[usenames]{color}
 \usepackage {tikz}
\usetikzlibrary {positioning}
\usepackage{mathtools}
\usepackage{xparse}
\DeclarePairedDelimiterX{\set}[1]{\{}{\}}{\setargs{#1}}
\NewDocumentCommand{\setargs}{>{\SplitArgument{1}{;}}m}
{\setargsaux#1}
\NewDocumentCommand{\setargsaux}{mm}
{\IfNoValueTF{#2}{#1} {#1\,\delimsize|\,\mathopen{}#2}}%{#1\:;\:#2}

\DeclarePairedDelimiter\abs{\lvert}{\rvert}

\DeclarePairedDelimiter\parenv{\lparen}{\rparen}

\theoremstyle{plain}
\newtheorem{theorem}{Theorem}
\newtheorem{lemma}[theorem]{Lemma}

\newtheorem{corollary}[theorem]{Corollary}
\newtheorem{definition}[theorem]{Definition}
\newtheorem{example}[theorem]{Example}
\newtheorem{remark}[theorem]{Remark}

\newcommand{\cC}{\mathcal{C}}

\renewcommand{\leq}{\leqslant}

\renewcommand{\geq}{\geqslant}

\newcommand{\N}{\mathbb{N}}
\newcommand{\R}{\mathbb{R}}

\DeclareMathOperator{\src}{src}
\DeclareMathOperator{\dest}{dest}

\DeclareMathOperator{\wt}{wt}

\newcommand{\Eout}{E_{\mathrm{out}}}
\newcommand{\Ein}{E_{\mathrm{in}}}
\newcommand{\DB}{De Bruijn }
\DeclareMathOperator*{\E}{\mathbb{E}}
\newcommand{\eqdef}{\triangleq}
\newcommand{\din}{d_{\mathrm{in}}}
\newcommand{\dout}{d_{\mathrm{out}}}
\newcommand{\te}{\tilde{e}}
\newcommand{\Cin}{C_{\mathrm{in}}}
\newcommand{\Cout}{C_{\mathrm{out}}}

\begin{document}

%----------------- The Title Declarations ------------------------------

\title{Improved Rank-Modulation Codes for DNA Storage with Shotgun Sequencing}

\author{\large
Niv Beeri and Moshe~Schwartz,~\IEEEmembership{Senior Member,~IEEE}%
\thanks{Niv Beeri is with the School
   of Electrical and Computer Engineering, Ben-Gurion University of the Negev,
   Beer Sheva 8410501, Israel
   (e-mail: nivbe@post.bgu.ac.il).}%
\thanks{Moshe Schwartz is with the School
   of Electrical and Computer Engineering, Ben-Gurion University of the Negev,
   Beer Sheva 8410501, Israel
   (e-mail: schwartz@ee.bgu.ac.il).}%
\thanks{This work was supported in part by the Israel Science Foundation (ISF) under grant No.~270/18.}
}

%% % make the title area
\maketitle

\begin{abstract}
  We study permutations over the set of $\ell$-grams, that are
  feasible in the sense that there is a sequence whose $\ell$-gram
  frequency has the same ranking as the permutation. Codes, which are
  sets of feasible permutations, protect information stored in DNA
  molecules using the rank-modulation scheme, and read using the
  shotgun sequencing technique. We construct systematic codes with an
  efficient encoding algorithm, and show that they are optimal in
  size. The length of the DNA sequences that correspond to the
  codewords is shown to be polynomial in the code
  parameters. Non-systematic with larger size are also constructed.
\end{abstract}


%%%%%%%%%%%%%%%%%%%%%%%%%%%%%%%%%%%%%%%%%%%%%%%%%%%%%%%%%%%%%%%
%%%%%%%%%%%%%%%%%%%%%%%%%%%%%%%%%%%%%%%%%%%%%%%%%%%%%%%%%%%%%%%
%%%%%%%%%%%%%%%%%%%%%%%%%%%%%%%%%%%%%%%%%%%%%%%%%%%%%%%%%%%%%%%

\section{Introduction}

\IEEEPARstart{S}{toring} information in DNA molecules offers
unparalleled information density, and has been proven to be feasible
\cite{BorLopCarDouCezSeeStr16,ChuGaoKos12,GolBerCheDesLepSipBri13,YazYuaMaZhaMil15}. Long
DNA sequences may be read relatively accurately using the \emph{shotgun
  sequencing} technique (see~\cite{MotBreTse12} and the references
therein). In this method, several copies of the same DNA sequence are
broken down into fragments. These fragments are identified, and an
algorithm reconstructs the DNA sequence using the knowledge of the
multiset of fragments obtained. Other similar variants of this
reconstruction method have also been
studied~\cite{GabOlg19,PatGabMil19,AchDasMilOrlPan15,GabPatMil20}.

It has been suggested by~\cite{KiaPulMil16} that we may skip the final
phase of sequence reconstruction, instead opting to have the
information encoded in the multiset of fragments. More precisely, if
the sequence is over an alphabet $\Sigma$, the shotgun sequencing
procedure provides us with a histogram, or a \emph{profile vector},
counting how many times each $\ell$-gram from $\Sigma^\ell$ appears as a
substring of the DNA sequence. Thus, the actual sequence is of no
consequence, acting merely a vehicle for its profile vector. As a side
benefit, this allows us to use ambiguous profile vectors that may
describe more than one sequence.

The profile vector obtained as part of the shotgun-sequencing
procedure is unfortunately noisy. Errors in it are mainly due to
substitution errors in the sequence-synthesis phase, non-uniform
fragmentation causing coverage gaps, and $\ell$-gram substitutions due
to sequencing (see~\cite{KiaPulMil16} and the references therein). One
approach, studied in~\cite{KiaPulMil16} is to protect the profile
vector using an error-correcting code, where an appropriate metric is
formulated to capture the error patterns mentioned.

Another suggestion put forth by~\cite{KiaPulMil16}, and later studied
by~\cite{RavSchYaa19}, was to employ the rank-modulation scheme over
the profile vectors. Rank modulation has a long history, starting
with~\cite{Sle65,BerJelWol72,ChaKur69} for vector digitization and
signal detection, through communication over power
lines~\cite{VinHaeWad00}, and more recently, for information storage
in non-volatile memories~\cite{JiaMatSchBru09}. In our context,
instead of storing the information in the profile vector, whose
integer entries count the number of occurrences of each $\ell$-gram from
$\Sigma^\ell$, the information is stored in the permutation over
$\Sigma^\ell$ which is the ranking (by frequency of appearance) of the
entries of the profile vector. By doing so we immediately gain a layer
of protection since perturbations of the profile vector that do not
result in a change of ranking, do not corrupt the stored
information. Additionally, there are known error-correcting codes for
the rank-modulation scheme, which we may use to gain further
protection~\cite{JiaSchBru10,TamSch10,BarMaz10,YehSch12b,MazBarZem13,ZhoSchJiaBru15,HorEtz14,ZhaGe16a,Hol17,ZhaGe16b,YehSch17}.

Not all permutations on $\Sigma^\ell$ correspond to a ranking of a
profile vector of some sequence, as was observed
in~\cite{RavSchYaa19}. A linear programming algorithm was derived
in~\cite{RavSchYaa19}, which can decide whether a given permutation is
feasible. However, an exact characterization of all feasible
permutations is still unknown. Thus, \cite{RavSchYaa19} provided only
upper bounds on the number of feasible permutations, and recursive
constructions that may also act as encoders. These constructions
produce codes whose rate is asymptotically $\frac{1}{\ell}$ when
$\ell$ is constant and the alphabet size $q=\abs{\Sigma}$ goes to
infinity, and $0$ when $q$ is fixed and $\ell\to\infty$. Additionally,
the length of the resulting encoded sequence was bounded and shown to
be polynomial in $q^\ell$. We also note that while~\cite{KiaPulMil16}
suggested the rank-modulation scheme, it did so only for a strict
subset of the entries of the profile vector.

The goal of this paper is to construct rank-modulation codes that
improve upon the best known ones, namely those
from~\cite{RavSchYaa19}.  Our main contributions are the following: We
construct systematic codes for all alphabet sizes $q\geq 3$, and all
window sizes $\ell\geq 2$. We give an efficient encoding algorithm for
these codes. The asymptotic rate of these codes is $1$ when $\ell$ is
fixed and $q\to\infty$, and is $1-\frac{1}{q}$ when $q$ is fixed and
$\ell\to\infty$, improving upon~\cite{RavSchYaa19}. The length of the
encoded sequence is analyzed and upper bounded by $O(q^{5\ell})$ for
$\ell\geq 3$, and $O(q^6)$ when $\ell=2$. These improve upon the order
of the corresponding bounds from~\cite{RavSchYaa19}. Additionally, our
upper bound is numerically lower than that of~\cite{RavSchYaa19}
except for the case of $q=3$ and $\ell=2$. We also prove an upper
bound on the size of systematic codes, which shows our construction
produces optimal systematic codes. Finally, we show a construction of
non-systematic codes that gives codes which are strictly larger than
their systematic counterparts.

The paper is organized as follows. In Section~\ref{sec:pre} we give
the necessary definitions used throughout the
paper. Section~\ref{sec:sys} we construct systematic codes, provide an
encoder, analyze the resulting sequence length, and prove an upper
bound on the size of such codes. In Section~\ref{sec:nonsys} we build
larger codes that are non-systematic. We conclude in
Section~\ref{sec:conc} with a summary and discussion of the results,
as well as some open problems.

%%%%%%%%%%%%%%%%%%%%%%%%%%%%%%%%%%%%%%%%%%%%%%%%%%%%%%%%%%%%%%%
%%%%%%%%%%%%%%%%%%%%%%%%%%%%%%%%%%%%%%%%%%%%%%%%%%%%%%%%%%%%%%%
%%%%%%%%%%%%%%%%%%%%%%%%%%%%%%%%%%%%%%%%%%%%%%%%%%%%%%%%%%%%%%%

\section{Preliminaries}
\label{sec:pre}

Throughout the paper we use $\Sigma$ to denote an alphabet of size
$q$. We assume no further structure on the alphabet. We use
$\Sigma^\ell$ to denote the set of all strings over $\Sigma$ of length
$\ell$, also called $\ell$-grams, and $\Sigma^*$ to denote the set of
all finite strings over $\Sigma$. If $s,s'\in\Sigma$ are strings, we
use $ss'$ to denote their concatenation, and $\abs{s}$ to denote the
length of $s$. If the need arises to consider specific letters in a
string $s\in\Sigma^n$, we shall usually denote the $i$th letter as
$s_i$, namely, $s=s_0 s_1 \dots s_{n-1}$, where $s_i\in\Sigma$ for all
$i\in[n]\eqdef\set{0,1,\dots,n-1}$.

If $G=(V,E)$ is a directed graph, we denote the edge $e\in E$ from
$v\in V$ to $v'\in V$ by $e=v\to v'$. We shall also say its source is
$\src(e)=v$ and its destination is $\dest(e)=v'$. Additionally, for
any vertex $v\in V$ we denote by $\Ein(v)$ the set of edges entering
$v$, and similarly, we use $\Eout(v)$ to denote the set of edges
leaving $v$, i.e.,
\begin{align*}
  \Ein(v) &\eqdef \set{ e\in E ; \dest(e)=v }, &
  \Eout(v) &\eqdef \set{ e\in E ; \src(e)=v }.
\end{align*}
The in-degree and out-degree of $v$ are similarly defined,
\begin{align*}
  \din(v) & \eqdef\abs*{\Ein(v)} & \dout(v)&\eqdef \abs*{\Eout(v)}.
\end{align*}
These definitions are extended to sets of vertices in the natural way.
Let $V'\subseteq V$ be a subset of vertices. Then we define
\begin{align*}
  \Ein(V') &\eqdef \set{ e\in E ; \dest(e)\in V', \src(e)\not\in V' }, &
  \Eout(V') &\eqdef \set{ e\in E ; \src(e)\in V', \dest(e)\not\in V' }.
\end{align*}

\subsection{Strings, profiles, and weighted De Bruijn graphs}

A useful tool in the context of string analysis is the De Bruijn
graph, which is defined as follows.
\begin{definition}
  The De Bruijn graph of order $\ell\geq 1$ over $\Sigma$ is the directed
  graph $G_{q,\ell}$ whose vertex set is $V(G_{q,\ell})=\Sigma^\ell$, and
  whose edge set is
  \[ E(G_{q,\ell})=\set*{ w_0 w_1 \dots w_{\ell-1} \to w_1 w_2 \dots w_\ell ;
    \text{for all $w_i\in\Sigma$}}.\]
\end{definition}

We observe that each edge $w_0 w_1 \dots w_{\ell-1} \to w_1 w_2 \dots
w_\ell$ in $G_{q,\ell}$ is uniquely identified by $w_0 \dots
w_\ell\in\Sigma^{\ell+1}$. Let $s=s_0\dots s_{n-1}\in\Sigma^n$ be a
string. We say that $s_i s_{i+1} \dots s_{i+\ell-1}$ is a window of
length $\ell$ into $s$, where indices are taken modulo $n$ (i.e., we
consider the string cyclically). Thus, by scanning $s$ with a sliding
window of length $\ell$, we obtain a cycle in $G_{q,\ell}$ whose
sequence of vertices corresponds to the windows into
$s$. Alternatively, with the same sliding window of length $\ell$ we
obtain a cycle in $G_{q,\ell-1}$ whose sequence of edges corresponds
to the windows into $s$. This latter correspondence between cycles in
$G_{q,\ell-1}$ and strings will be used throughout the paper.

Motivated by the process of shotgun sequencing, previous
papers~\cite{KiaPulMil16,RavSchYaa19} suggested that information be
encoded in the profile vector of the DNA sequence, whose definition
follows.

\begin{definition}
  Let $s=s_0 \dots s_{n-1}\in\Sigma^n$ be a string. The \emph{profile
    vector of $s$ of order $\ell$}, denoted by
  $p_{s,\ell}\in(\N\cup\set{0})^{\Sigma^\ell}$, is a non-negative
  integer vector indexed by $\Sigma^\ell$ such that for each
  $w\in\Sigma^\ell$,
  \[ p_{s,\ell}(w)\eqdef \abs*{\set*{ i\in[n] ; s_i s_{i+1} \dots s_{i+\ell-1}=w}},\]
  where indices are taken modulo $n$. Namely, $p_{s,\ell}(w)$ counts the
  number of occurrences of $w$ in $s$ (cyclically).
\end{definition}

\begin{definition}
  Let $x\in(\N\cup\set{0})^{\Sigma^\ell}$. We say $x$ is
  \emph{feasible} if there exists $s\in\Sigma^*$ whose profile vector
  of order $\ell$ is $x$, namely, $p_{s,\ell}=x$.
\end{definition}

\begin{example}
\label{ex:first}
Let $\Sigma = \set{A,C,G}$, and consider the string
\[s = GGGGAGAGAGGGGAAAAAAAACCCCCCCAGGGGCGCGCGCGCGCGCCCCAGCCGCCG.\]
The profile vector of $s$ of order $2$ is
\begin{equation}
  \label{eq:exfirst}
  p_{s,2} = (7,1,5,2,11,8,4,9,10),
\end{equation}
where the indices of the profile vector are in lexicographic order,
i.e., $AA,AC,AG,CA,CC,CG,GA,GC,GG$.
\end{example}

Not every vector $x\in(\N\cup\set{0})^{\Sigma^\ell}$ is feasible. Let
us build the following directed graph, $G$, with vertices
$V=\Sigma^{\ell-1}$, and for every $w=w_0\dots
w_{\ell-1}\in\Sigma^\ell$ we place $x(w)$ parallel copies of the edge
$w_0\dots w_{\ell-2}\to w_1\dots w_{\ell-1}$. Then by our previous
discussion of De Bruijn graphs, it is obvious that $x$ is the profile
vector of order $\ell$ of some string $s$ if and only if $G$ contains
an Eulerian cycle (i.e., a cycle passing through each edge exactly
once). In turn, an Eulerian cycle exists if and only if $G$ is
strongly connected (excluding isolated vertices) and for every vertex
$v\in V$, its in-degree equals its out-degree, $\din(v)=\dout(v)$.

For our convenience, we replace the $x(w)$ parallel edges discussed
above with a single edge of weight $x(w)$. In general, for a directed
graph $G=(V,E)$ we use $\wt_G(e)\in\R$ to denote the weight of an edge
$e\in E$. We omit the subscript $G$ if it is clear from context. We
also extend this definition to subsets of edges $E'\subseteq E$ by
defining $\wt(E')\eqdef \sum_{e\in E'}\wt(e)$. 

\begin{definition}
  Let $G=(V,E)$ be a directed weighted graph. We say $G$ is
  \emph{balanced} if $\wt(\Ein(v))=\wt(\Eout(v))$ for all $v\in V$.
\end{definition}

We therefore have the following corollary, translating our previous
observation that uses parallel edges, to one using weights.

\begin{lemma}
  A vector $x\in\N^{\Sigma^\ell}$ is feasible if and only if the
  weighted De Bruijn graph, $G_{q,\ell-1}$, with weights $\wt(e)=x(e)$
  for all $e\in\Sigma^\ell$, is balanced.
\end{lemma}
\begin{IEEEproof}
  Replace each edge $e$ with $\wt(e)$ parallel edges. Since the weight
  of every edge is positive, the resulting graph, $G'$, is strongly
  connected, and therefore $x$ is feasible if and only if
  $\din(v)=\dout(v)$ for every $v\in G'$. But that happens if and only
  if $G_{q,\ell-1}$ is balanced.
\end{IEEEproof}

Following~\cite{RavSchYaa19}, we shall almost always consider strings
$s$ whose profile vectors are all positive integers, i.e., for all
$w\in\Sigma^\ell$, $p_{s,\ell}(w)>0$.

\begin{figure}[t]
  \definecolor {processblue}{cmyk}{0.3,0.3,0.3,0.3}
  \begin {center}
    \begin {tikzpicture}[-latex ,auto ,node distance =4 cm and 3cm ,on grid ,
        semithick ,
        state/.style ={ circle ,top color =processblue!20 , bottom color = processblue!20 ,
          draw,processblue , text=black , minimum width =1.4 cm}]
      \node[state] (G){$\textbf{G}$};
      \node[state] (A) [above left=of G] {$\textbf{A}$};
      \node[state] (C) [above right =of G] {$\textbf{C}$};
      \path (A) edge [loop left] node[above = 0.15] {$7$} (A);
      \path (C) edge [bend left =15] node[below =0.05 cm] {$2$} (A);
      \path (A) edge [bend right = -15] node[below =0.15 cm] {$1$} (C);
      \path (A) edge [bend left =15] node[above] {$5$} (G);
      \path (G) edge [bend left =15] node[below =0.15 cm] {$4$} (A);
      \path (C) edge [bend left =15] node[below =0.15 cm] {$8$} (G);
      \path (G) edge [bend right = -15] node[below =0.15 cm] {$9$} (C);
      \path (G) edge [loop below] node[left = 0.2] {$10$} (G);
      \path (C) edge [loop right] node[above = 0.15] {$11$} (C);
    \end{tikzpicture}
  \end{center}
  \caption{The weighted De Bruijn graph of Example~\ref{ex:dbex1}. }
  \label{fig:dbex1}
\end{figure}
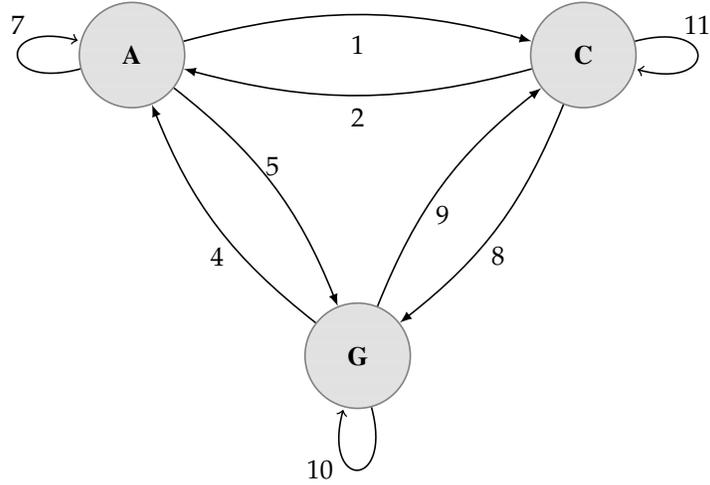

\begin{example}
  \label{ex:dbex1}
  We continue the setting of Example \ref{ex:first}. In
  Figure~\ref{fig:dbex1} we draw the De Bruijn graph with edge weights
  in accordance with the profile vector $p_{s,2}$
  of~\eqref{eq:exfirst}. We observe that the resulting graph is
  balanced, not surprising as the profile vector was taken from a
  string, i.e, the profile vector is feasible.
\end{example}

We would like to make one more simple observation that will be useful
later.

\begin{lemma}
  \label{lem:balset}
  Let $G=(V,E)$ be a finite weighted directed graph. Then $G$ is
  balanced if and only if for every $U\subseteq V$,
  $\wt(\Ein(U))=\wt(\Eout(U))$.
\end{lemma}

\begin{IEEEproof}
  One direction is trivial. If $\wt(\Ein(U))=\wt(\Eout(U))$ for all
  $U\subseteq V$, then it is true in particular for subsets $U$
  containing exactly one vertex, making $G$ balanced by definition.

  In the other direction, for any $U\subseteq V$ we have
  \[
  \wt(\Ein(U))-\wt(\Eout(U)) = \sum_{e\in\Ein(U)}\wt(e)-\sum_{e\in\Eout(U)}\wt(e)
  \overset{(a)}{=}\sum_{v\in U}\parenv{\wt(\Ein(v))-\wt(\Eout(v))} = 0,
  \]
  where $(a)$ follows from the fact that edges $v'\to v''$, where
  $v',v''\in U$, that are added to the sum $\sum_{v\in
    U}\wt(\Ein(v))$, are also added to the sum $\sum_{v\in
    U}\wt(\Eout(v))$, thus, canceling out.
\end{IEEEproof}

\subsection{Permutations and rank modulation}

Let $A$ be a finite set. We use $S_A$ to denote the set of
permutations over $A$. Each permutation $\pi\in S_A$ may be considered
as a bijection $A\to [\abs{A}]$, sending each element of $A$ to its
unique ranking in the permutation. Encoding information in
permutations of the set $A$, instead of vectors over $A$, has a long
history under the name \emph{rank modulation}. The identity of the set
$A$ depends on the specifics of the applications. As examples we
bring~\cite{ChaKur69} dealing with signal detection with impulsive
noise, ~\cite{VinHaeWad00} for powerline communications,
and~\cite{JiaMatSchBru09} for coding in flash memories.

Recently,~\cite{KiaPulMil16} suggested applying the rank-modulation
scheme to DNA storage, with a follow-up
work~\cite{RavSchYaa19}. There, the set of permutations is
$S_{q,\ell}\eqdef S_{\Sigma^\ell}$, and the ranking is done by the
entries of the profile vector of the DNA sequence. Precise definitions
follow:

\begin{definition}
  Let $\pi\in S_{q,\ell}$ be a permutation, and let
  $x\in\N^{\Sigma^\ell}$ be some vector. We say that $x$ satisfies
  $\pi$, writing $x\vDash\pi$, if the entries of $x$ are distinct and
  and for $w,w'\in\Sigma^\ell$, $\pi(w)<\pi(w')$ if and only if
  $x(w)<x(w')$. Additionally, we say $\pi$ is feasible if $x$ is
  feasible.
\end{definition}

We denote the set of all feasible permutations over $\Sigma^\ell$ by
$\Phi_{q,\ell}$, and their number by
$F_{q,\ell}\eqdef\abs*{\Phi_{q,\ell}}$. Since we will also be
interested in rates, in the coding-theoretic meaning, for any
non-empty subset $\cC\subseteq S_A$, we define its \emph{rate} as
\[ R(\cC)\eqdef \frac{\log_2\abs*{\cC}}{\log_2\abs*{S_A}}.\]
We can then define the feasible rate as
\[ R_{q,\ell} \triangleq R(\Phi_{q,\ell})=\frac{\log_2 \abs*{\Phi_{q,\ell}}}{\log_2 \abs*{S_{q,\ell}}}=\frac{\log_2F_{q,\ell}}{\log_2(q^\ell
  !)}.\]

\begin{example}
  \label{ex:per}
  We continue Example~\ref{ex:dbex1}. The profile vector $p_{s,2}$
  of~\eqref{eq:exfirst} induces a feasible permutation $\pi \in
  S_{3,2}$ as follows:
  \begin{equation}
    \label{eq:exper}
    \pi=\begin{pmatrix}
    AA & AC & AG & CA & CC & CG & GA & GC & GG \\
    4  & 0  & 3  & 1  & 8  & 5  & 2  & 6  & 7 \end{pmatrix},
  \end{equation}
  presented in the standard two-line notation.
\end{example}

We shall need the following projection operator for permutations.

\begin{definition}
  Let $B\subseteq A$ be two finite sets, and let $\pi\in S_A$ be a
  permutation over $A$. We use $\pi|_B$ to denote the unique
  permutation in $S_B$ that keeps the relative order of the elements
  of $B$ in $\pi$, namely, for all $b,b'\in B$, $\pi|_B(b)<\pi|_B(b')$
  if and only if $\pi(b)<\pi(b')$.
\end{definition}

We can think of $\pi|_B$ in the previous definition as the projection
of $\pi$ onto the elements in the set $B$.

\begin{example}
  We continue Example~\ref{ex:per}. Taking the permutation $\pi$
  of~\eqref{eq:exper}, for the set $B = \set{AC,CC,GA,GC}$ we have
  \[\pi|_B = \begin{pmatrix}
    AC & CC & GA & GC \\
    0  & 3  & 1  & 2 \end{pmatrix},\]
  since $\pi(AC)<\pi(GA)<\pi(GC)<\pi(CC)$.
\end{example}

As usual in rank modulation, we define a code $\cC$ to be a subset of
$S_A$. If $\abs{A}=n$ and $\abs{\cC}=M$, we say that $\cC$ is an
$(n,M)$-code. Of particular interest to us are systematic codes, which
are analogous to systematic linear codes.

\begin{definition}
  Let $A$ be some set, $\abs{A}=n$. We say $\cC\subseteq S_A$ is an
  \emph{$[n,k]$-systematic code}, if there exists a set $B\subseteq
  A$, $\abs{B}=k$, $\abs{\cC}=k!$, and
  \[ \set*{\pi|_B ; \pi\in\cC} = S_B.\]
  We call $B$ an \emph{information set} for the code $\cC$.
\end{definition}

Intuitively, in a systematic code the user may set the ranking of the
information set, $B\subseteq A$, arbitrarily (thereby, storing the
user information). The remaining entries of the permutation,
$A\setminus B$, are then determined by the code, creating a
permutation over $A$.

\section{Optimal Systematic Codes for Feasible Permutations}
\label{sec:sys}

In this section we study systematic codes for feasible permutation,
namely, systematic subsets $\cC\subseteq \Phi_{q,\ell}$. We provide a
construction for such codes for all parameters, and show an efficient
encoding algorithm. We further prove these are optimal, i.e., having
the largest possible size of all systematic codes. Additionally, we
analyze the length of the realizing strings, and show they are at most
polynomial in the trivial lower bound.

\subsection{Construction}

We start by giving some technical lemmas. The first shows two basic
operations that take a balanced directed graph, modify the weights,
but keep it balanced.

\begin{lemma}
  \label{lem:keepbal}
  Let $G=(V,E)$ be a finite balanced directed graph. Construct $G'=(V,E)$. Then:
  \begin{enumerate}
  \item
    If for all $e\in E$, $\wt_{G'}(e)=c\cdot\wt_{G}(e)$, where $c\in\R$
    is some constant, then $G'$ is also balanced.
  \item
    Let $e_0,e_1,\dots,e_{m-1}$ be a sequence of edges in $E$ that form a
    cycle, and let $c\in\R$ be some constant. If
    \[ \wt_{G'}(e)=\begin{cases}
    \wt_{G}(e)+c & \text{$e=e_i$ for some $i$,} \\
    \wt_{G}(e) & \text{otherwise,}
    \end{cases}\]
    then $G'$ is also balanced.
  \end{enumerate}
\end{lemma}
\begin{IEEEproof}
  Multiplying the weights by a constant naturally keeps all vertices
  balanced. For the second case, we note that vertices that reside on
  the cycle have the same number of edges from the cycle entering as
  there are leaving. Thus, adding a constant weight to the edges of
  the cycle keeps the graph balanced.
\end{IEEEproof}

Another simple lemma states that if we know that all but one of the
vertices are balanced, then that vertex is also balanced.

\begin{lemma}
\label{lem:bal}
Let $G=(V,E)$ be a directed weighted graph, and let $v\in V$ be some
vertex. If $\wt(\Ein(v')) = \wt(\Eout(v'))$ for all $v'\in
V\setminus\set{v}$, then also $\wt(\Ein(v))=\wt(\Eout(v))$.
\end{lemma} 

\begin{IEEEproof}
  We observe that $\set{ \Ein(v') ; v'\in V}$ is a partition of $E$, as is
  $\set{\Eout(v') ; v'\in V}$. Thus,
  \[ \wt(\Ein(v)) = \wt(E)-\sum_{v'\in V\setminus\set{v}} \wt(\Ein(v'))
  = \wt(E)-\sum_{v'\in V\setminus\set{v}} \wt(\Eout(v')) = \wt(\Eout(v)),\]
  which proves the claim.
\end{IEEEproof}

We recall that a Hamiltonian cycle/path in a graph visits every vertex
exactly once, whereas an Eulerian cycle/path visits every edge exactly
once. It is well known (see~\cite{VanDeb51}) that a Hamiltonian cycle
in the De Bruijn graph $G_{q,\ell}$ (which is equivalent to a De
Bruijn sequence) exists for all $q,\ell\geq 2$. Such a
Hamiltonian cycle is also equivalent to an Eulerian cycle in
$G_{q,\ell-1}$.

De Bruijn sequences may be nested, with lower-order sequences being
prefixes of higher-order sequences. We cite the following result
from~\cite{BecAri11}.

\begin{lemma}
\label{lem:HtE}
\cite[Th.~1]{BecAri11} Let $G_{q,\ell-1}$ be a \DB graph with $q\geq
3$ and $\ell\geq 2$, then every Hamiltonian cycle in $G_{q,\ell-1}$
can be extended to an Eulerian cycle.
\end{lemma}

We shall further need the following technical lemma, which shows that
we can complete cycles while avoiding a given Hamiltonian path.

\begin{lemma}
  \label{lem:cyc}
  Let $G_{q,\ell-1}$ be a \DB graph, $q\geq 3$, $\ell\geq 2$, and let
  $E_H=\set{e_0,e_1,\dots,e_{q^{\ell-1}-1}}$ be the edges of a
  Hamiltonian cycle in $G_{q,\ell-1}$. Then for any
  $i\in[q^{\ell-1}]$, there is a cycle passing through $e_i$ while not
  passing through any $e_j$, $j\neq i$.
\end{lemma}
 
\begin{IEEEproof}
  As a consequence of Lemma~\ref{lem:HtE}, after removing the edges of
  $E_H$ from $G_{q,\ell-1}$, there exists an Eulerian cycle, $\beta$,
  in the remaining graph.  Let $e_i=v_i\to v_{i+1}$ (where indices are
  taken modulo $q^{\ell-1}$) be the edge that we wish to complete to
  a cycle. Since $q\geq 3$, there exists at least one outgoing edge
  from $v_{i+1}$ and one incoming edge to $v_{i}$ that are not in
  $E_H$. Thus, at some point $\beta$ leaves $v_{i+1}$ and at some
  point it enters $v_i$. Denote by $\tilde{\beta}$ a part of $\beta$
  that forms a path $v_{i+1}\leadsto v_i$.  It now follows that
  $e_i,\tilde{\beta}$ is a cycle passing through $e_i$ but avoiding
  all $e_j$, $j\neq i$.
\end{IEEEproof}

We are now ready for the main theorem of this section, that construct
a large systematic code in the space of feasible permutations.

\begin{theorem}
  \label{th:inject}
  Let $G_{q,\ell-1}=(V,E)$ be a \DB graph, $q\geq 3$, $\ell\geq
  2$. Let $e_0,e_1,\dots,e_{q^{\ell-1}-1}$ be a sequence of edges
  forming a Hamiltonian cycle in $G_{q,\ell-1}$, and define
  $E'_H\eqdef \set{e_0,\dots,e_{q^{\ell-1}-2}}$.  Then there is an
  injective mapping $S_{E\setminus E'_H}\to \Phi_{q,\ell}$, namely,
  between the set of permutations over $E\setminus E'_H$ and the set
  of feasible permutations over $\Sigma^{\ell}$.
\end{theorem}

\begin{IEEEproof}
Let us index the vertices of $G_{q,\ell-1}$ as $v_0,v_1,\dots
,v_{q^{\ell-1}-1}$, ordered such that $e_i=v_i\to v_{i+1}$ for all
$i\in[q^{\ell-1}]$ (and where indices are taken modulo
$q^{\ell-1}$). We need to show that for every permutation on
$E\setminus E'_H$ we can build a distinct feasible permutation on
$\Sigma^{\ell} \cong E$.

Let $\pi\in S_{E\setminus E'_H}$ be any permutation on $E\setminus
E'_H$. We assign the edges in $E\setminus E'_H$ distinct positive
weights while keeping the ranking as in $\pi$. This is easily achieved
by setting $\wt(e)=\pi(e)+1$ for all $e\in E\setminus E'_H$. Notice
that because the edges in $E'_H$ form a Hamiltonian path, then every
vertex in $V\setminus\set{v_{q^{\ell-1}-1}}$ is left with exactly one
outgoing edge whose weight has not been set yet.

In the next step, we assign weights to the remaining edges, i.e., the
edges in $E'_H$. We do so in such a way that all vertices become
balanced. For $i = 0,1,\dots,{q^{\ell-1}-2}$, in that order, we assign
the weight of $e_i$ to be
\[ \wt(e_i) = \wt(\Ein(v_i)) - \wt(\Eout(v_i)\setminus\set{e_i}).\]
Thus, all the vertices in $V\setminus\set{v_{q^{\ell-1}-1}}$ are
balanced. By Lemma~\ref{lem:bal} we must have that $v_{q^{\ell-1}-1}$ is
also balanced.

At this point we have assigned integer weights to all of the edges. In
order for the weights to induce a permutation over $E$, we need them
to be distinct. This is certainly true, by construction, for the edges
in $E\setminus E'_H$. However, following the balancing process that
set the weights for edges in $E'_H$, we are not guaranteed
distinctness of weights for edges in $E$, and we therefore need to
break ties. For the remainder of the proof we proceed with slightly
different sets of edges. Define $E_H\eqdef
E'_H\cup\set{e_{q^{\ell-1}-1}}$ to be the set of edges in the
Hamiltonian cycle required by the theorem. Since $E\setminus E_H
\subseteq E\setminus E'_H$, the weights of edges in $E\setminus E_H$
are distinct. It follows that there are only two cases in which we
could be seeing equality between weights of two edges: the two edges
are from $E_H$, or one edge is from $E_H$ and the other from
$E\setminus E_H$.

Let us start by resolving the first case. For each
$i\in[q^{\ell-1}-1]$, let $\gamma_i$ be a cycle in
$G_{q,\ell-1}$ that contains $e_i\in E_H$ but does not contain any
$e\in E_H$, $e\neq e_i$. The existence of such cycles is guaranteed by
Lemma~\ref{lem:cyc}. We define
\[ \Delta \eqdef \binom{q^{\ell-1}}{2}+1=\frac{q^{\ell-1}(q^{\ell-1}-1)}{2}+1,\]
and then add $(i+1)\cdot\Delta^{-1}$ to the weight of each of the edges in
$\gamma_i$, for all $i\in[q^{\ell-1}-1]$. By
Lemma~\ref{lem:keepbal}, we therefore keep the graph balanced. We
further observe that the maximum total weight added to any single edge
is upper bounded by
\begin{equation}
  \label{eq:delta}
  \sum_{i=0}^{q^{\ell-1}-2}\frac{i+1}{\Delta}=\frac{1}{\Delta}\cdot\binom{q^{\ell-1}}{2}\leq 1-\frac{1}{\Delta}.
\end{equation}
It follows that if $\wt(e)>\wt(e')$ before the weight addition, then
this relation remains unchanged after the weight addition. In
particular, the ranking of edges by weight in $E\setminus E'_H$
remains unchanged. Additionally, since distinct weights in the
interval $[0,1-\Delta^{-1}]$ were added to the integer weights of
edges of $E_H$, all weights of edges in $E_H$ are now distinct.

For the second case, we increase the weights of edges in the
Hamiltonian cycle, $E_H$, by $\frac{1}{2}\Delta^{-1}$. By
Lemma~\ref{lem:keepbal}, the graph remains balanced. However, now
edges in $E\setminus E_H$ have weights that are integer multiples of
$\Delta^{-1}$, whereas the weights of edges in $E_H$ are not.
Additionally, by \eqref{eq:delta}, the addition of
$\frac{1}{2}\Delta^{-1}$ to the weight does not change the ranking of
edges in $E\setminus E'_H$. Thus, the second case is resolved as well.

We are now in possession of a weighted graph that is balanced, while
keeping the relative ranking of weights of edges in $E\setminus E'_H$,
and having distinct weights. Using Lemma~\ref{lem:keepbal}, we now
multiply all the edge weights by $2\Delta$, to obtain the same
properties mentioned above, only with integer weights. Finally, we
subtract $\min_{e\in E}\wt(e)-1$ from all the weights. Since
$G_{q,\ell-1}$ is Eulerian, by Lemma~\ref{lem:keepbal}, the resulting
weights are positive integers, and the graph has the properties
mentioned above.

Let us denote the permutation induced by the weights of the
edges by $\pi'\in S_E$. Clearly, by the previous discussion,
\[ \pi'|_{E\setminus E'_H} = \pi,\]
hence the mapping described here, $S_{E\setminus E'_H}\to S_E$ is
injective.  Furthermore, since the resulting weighted graphs are all
balanced, all resulting permutations are feasible and this mapping is
in fact $S_{E\setminus E'_H}\to \Phi_{q,\ell}$.
\end{IEEEproof}

The algorithm described in the proof of Theorem~\ref{th:inject} is
summarized using pseudocode as Algorithm~\ref{alg:A}.

\begin{algorithm}[t]
  \DontPrintSemicolon
  \SetKwInOut{Input}{Input}
  \SetKwInOut{Output}{Output}
  \SetKwInOut{Parameters}{Parameters}
  \Parameters{%
    Alphabet $\Sigma$, $\abs{\Sigma}=q\geq 3$, $\ell\geq 2$, the De Bruijn graph $G_{q,\ell-1}=(V,E)$\\
    A sequence of edges $\alpha=e_0,\dots,e_{q^{\ell-1}-1}$ forming a Hamiltonian cycle in $G_{q,\ell-1}$,\\
    A sequence of edges $\beta=\te_0,\dots,\te_{q^{\ell}-q^{\ell-1}-1}$ such that $\alpha,\beta$ is an Eulerian cycle in $G_{q,\ell-1}$.
  }
  \Input{%
    A permutation $\pi\in S_{E\setminus E'_H}$, where $E'_H\eqdef\set{e_0,\dots,e_{q^{\ell-1}-2}}$.
  }
  \Output{%
    A profile vector $x\in\N^{\Sigma^\ell}$ such that $x\vDash \pi'\in S_E$ and $\pi'|_{E\setminus E'_H}=\pi$.
  }
  \tcp{Initial systematic part values}
  \For{$i\gets 0$ \KwTo $q^{\ell}-q^{\ell-1}-1$}{
    $x(\te_i) \gets \pi(\te_i)+1$
  }
  $x(e_{q^{\ell-1}-1})\gets \pi(e_{q^{\ell-1}-1})+1$\\
  \tcp{Balance the graph}
  \For{$i\gets 0$ \KwTo $q^{\ell-1}-2$}{
    $x(e_i) \gets \sum_{e\in\Ein(\src(e_i))}x(e)-\sum_{e\in\Eout(\src(e_i))} x(e)$
  }
  \tcp{Break ties in $\alpha$}
  $\Delta\gets \binom{q^{\ell-1}}{2}+1$\\
  \For{$i\gets 0$ \KwTo $q^{\ell-1}-2$}{
    Find $s,s'$ such that $\src(e_i)=\dest(\te_{s'})$ and $\dest(e_i)=\src(\te_s)$\\
    \For{$j\gets s$ \KwTo $s'$ (cyclically)}{
      $x(\te_j)\gets x(\te_j)+(i+1)\cdot \Delta^{-1}$
    }
    $x(e_i)\gets x(e_i)+(i+1)\cdot\Delta^{-1}$
  }
  \tcp{Break ties between $\alpha$ and $\beta$}
  \For{$i\gets 0$ \KwTo $q^{\ell-1}-1$}{
    $x(e_i)\gets x(e_i)+\frac{1}{2}\Delta^{-1}$
  }
  \tcp{Make weights integers}
  \ForAll{$e\in E$}{
    $x(e)\gets 2\Delta\cdot x(e)$
  }
  \tcp{Make weights start at $1$}
  $m\gets \min_{e\in E} x(e)$\\
  \ForAll{$e\in E$}{
    $x(e)\gets x(e)-m+1$
  }
  \caption{A systematic encoding algorithm for any $q\geq 3$ and $\ell\geq 2$}
  \label{alg:A}
\end{algorithm}

\begin{figure*}[t]
  \centering
  \definecolor {processblue}{cmyk}{0.3,0.3,0.3,0.3}
  \subfloat[]{
    \begin {tikzpicture}[-latex ,auto ,node distance =4 cm and 4.5cm ,on grid ,
        semithick ,
        state/.style ={ circle ,top color =processblue!20 , bottom color = processblue!20 ,
          draw,processblue , text=black , minimum width =1.4 cm}]
      \node[state] (G){$\textbf{G}$};
      \node[state] (A) [above =of G] {$\textbf{A}$};
      \node[state] (T) [right =of G] {$\textbf{T}$};
      \node[state] (C) [above =of T] {$\textbf{C}$};
      \path (A) edge [loop left] node[above = 0.15] {$10$} (A);
      \path (G) edge [bend left =15] node[above left= 0.04 cm]{$9$} (A);
      \path (A) edge [dashed][bend right = -15] node[below left =0.04 cm] {$ $} (G);
      \path (A) edge  [bend left=11, align=left, sloped, inner sep=1pt] node[right = 1,above ] {$ 7$} (T);
      \path (T) edge [bend left=11, align=left, sloped, inner sep=1pt] node[left = 1,below] {$6$} (A);
      \path (G) edge [dashed][bend left =15] node[below] {$ $} (T);
      \path (T) edge [bend left =12] node[below ] {$8$} (G);
      \path (T) edge [loop right] node[above = 0.15] {$12$} (T);
      \path (G) edge [loop left ] node[above = 0.15] {$11$} (G);
      \path (C) edge [loop right] node[above = 0.15] {$13$} (C);
      \path (C) edge [bend right = -15] node[above right =0.04 cm] {$3$} (T);
      \path (A) edge [bend left =15] node[above] {$1$} (C);
      \path (C) edge [bend left =12] node[above] {$2$} (A);
      \path (T) edge [dashed][bend left =15] node[below right= 0.04 cm] {$ $} (C);
      \path (C) edge [bend left=11, align=left, sloped, inner sep=1pt] node[left= 1,below ] {$5$} (G);
      \path (G) edge [bend left=11, align=left, sloped, inner sep=1pt] node[right= 1,above ] {$4$} (C);
    \end{tikzpicture}
    \label{fig:case:1}
  }
  \hfil
  \subfloat[]{
    \begin {tikzpicture}[-latex ,auto ,node distance =4 cm and 4.5cm ,on grid ,
        semithick ,
        state/.style ={ circle ,top color =processblue!20 , bottom color = processblue!20 ,
          draw,processblue , text=black , minimum width =1.4 cm}]
      \node[state] (G){$\textbf{G}$};
      \node[state] (A) [above =of G] {$\textbf{A}$};
      \node[state] (T) [right =of G] {$\textbf{T}$};
      \node[state] (C) [above =of T] {$\textbf{C}$};
      \path (A) edge [loop left] node[above = 0.15] {$10$} (A);
      \path (G) edge [bend left =15] node[above left= 0.04 cm]{$9$} (A);
      \path (A) edge [dashed][bend right = -15] node[below left =0.04 cm] {$\textbf{9}$} (G);
      \path (A) edge  [bend left=11, align=left, sloped, inner sep=1pt] node[right = 1,above ] {$ 7$} (T);
      \path (T) edge [bend left=11, align=left, sloped, inner sep=1pt] node[left = 1,below] {$6$} (A);
      \path (G) edge [dashed][bend left =15] node[below] {$\textbf{9}$} (T);
      \path (T) edge [bend left =12] node[below ] {$8$} (G);
      \path (T) edge [loop right] node[above = 0.15] {$12$} (T);
      \path (G) edge [loop left ] node[above = 0.15] {$11$} (G);
      \path (C) edge [loop right] node[above = 0.15] {$13$} (C);
      \path (C) edge [bend right = -15] node[above right =0.04 cm] {$3$} (T);
      \path (A) edge [bend left =15] node[above] {$1$} (C);
      \path (C) edge [bend left =12] node[above] {$2$} (A);
      \path (T) edge [dashed][bend left =15] node[below right= 0.04 cm] {$\textbf{5}$} (C);
      \path (C) edge [bend left=11, align=left, sloped, inner sep=1pt] node[left= 1,below ] {$5$} (G);
      \path (G) edge [bend left=11, align=left, sloped, inner sep=1pt] node[right= 1,above ] {$4$} (C);
    \end{tikzpicture}
    \label{fig:case:2}
  } \\
  \subfloat[]{
    \begin {tikzpicture}[-latex ,auto ,node distance =4 cm and 4.5cm ,on grid ,
        semithick ,
        state/.style ={ circle ,top color =processblue!20 , bottom color = processblue!20 ,
          draw,processblue , text=black , minimum width =1.4 cm}]
      \node[state] (G){$\textbf{G}$};
      \node[state] (A) [above =of G] {$\textbf{A}$};
      \node[state] (T) [right =of G] {$\textbf{T}$};
      \node[state] (C) [above =of T] {$\textbf{C}$};
      \path (A) edge [loop left] node[above = 0.15] {$10$} (A);
      \path (G) edge [bend left =15] node[above left= 0.04 cm]{$9+\frac{2}{2\Delta}$} (A);
      \path (A) edge [dashed][bend right = -15] node[below left =0.02 cm] {$\textbf{9}+\frac{2+1}{2\Delta}$} (G);
      \path (A) edge  [bend left=11, align=left, sloped, inner sep=1pt] node[right = 1,above ] {$ 7+\frac{4}{2\Delta}$} (T);
      \path (T) edge [bend left=11, align=left, sloped, inner sep=1pt] node[left = 1,below] {$6+\frac{4}{2\Delta}$} (A);
      \path (G) edge [dashed][bend left =15] node[below] {$\textbf{9}+\frac{4+1}{2\Delta}$} (T);
      \path (T) edge [bend left =12] node[below ] {$8+\frac{4}{2\Delta}$} (G);
      \path (T) edge [loop right] node[above = 0.15] {$12+\frac{4}{2\Delta}$} (T);
      \path (G) edge [loop left ] node[above = 0.15] {$11+\frac{2}{2\Delta}$} (G);
      \path (C) edge [loop right] node[above = 0.15] {$13+\frac{6}{2\Delta}$} (C);
      \path (C) edge [bend right = -15] node[above right =0.04 cm] {$3+\frac{6}{2\Delta}$} (T);
      \path (A) edge [bend left =15] node[above] {$1$} (C);
      \path (C) edge [bend left =12] node[above] {$2+\frac{1}{2\Delta}$} (A);
      \path (T) edge [dashed][bend left =15] node[below right= 0.04 cm] {$\textbf{5}+\frac{6+1}{2\Delta}$} (C);
      \path (C) edge [bend left=11, align=left, sloped, inner sep=1pt] node[left= 1,below ] {$5+\frac{2}{2\Delta}$} (G);
      \path (G) edge [bend left=11, align=left, sloped, inner sep=1pt] node[right= 1,above ] {$4+\frac{2}{2\Delta}$} (C);
    \end{tikzpicture}
    \label{fig:case:3}
  }
  \hfil
  \subfloat[]{
    \begin {tikzpicture}[-latex ,auto ,node distance =4 cm and 4.5cm ,on grid ,
        semithick ,
        state/.style ={ circle ,top color =processblue!20 , bottom color = processblue!20 ,
          draw,processblue , text=black , minimum width =1.4 cm}]
      \node[state] (G){$\textbf{G}$};
      \node[state] (A) [above =of G] {$\textbf{A}$};
      \node[state] (T) [right =of G] {$\textbf{T}$};
      \node[state] (C) [above =of T] {$\textbf{C}$};
      \path (A) edge [loop left] node[above = 0.15] {$127$} (A);
      \path (G) edge [bend left =15] node[above left= 0.04 cm]{$115$} (A);
      \path (A) edge [dashed][bend right = -15] node[below left =0.001 cm] {$\textbf{116}$} (G);
      \path (A) edge [bend left=11, align=left, sloped, inner sep=1pt] node[right = 1,above ] {$89$} (T);
      \path (T) edge [bend left=11, align=left, sloped, inner sep=1pt] node[left = 1,below]{$75$} (A);
      \path (G) edge [dashed][bend left =15] node[below] {$\textbf{118}$} (T);
      \path (T) edge [bend left =12] node[below ] {$103$} (G);
      \path (T) edge [loop right] node[above = 0.15] {$159$} (T);
      \path (G) edge [loop left ] node[above = 0.15] {$143$} (G);
      \path (C) edge [loop right] node[above = 0.15] {$175$} (C);
      \path (C) edge [bend right = -15] node[above right =0.04 cm] {$35$} (T);
      \path (A) edge [bend left =15] node[above] {$1$} (C);
      \path (C) edge [bend left =12] node[above] {$16$} (A);
      \path (T) edge [dashed][bend left =15] node[below right= 0.04 cm] {$\textbf{64}$} (C);
      \path (C) edge [bend left=11, align=left, sloped, inner sep=1pt] node[left= 1,below ] {$59$} (G);
      \path (G) edge[bend left=11, align=left, sloped, inner sep=1pt] node[right= 1,above ] {$45$} (C);
    \end{tikzpicture}
    \label{fig:case:4}
  }
  \caption{A depiction of Algorithm~\ref{alg:A} in the setting of
    Example~\ref{ex:alg}: (a) the user information weights, (b) the initial balancing, (c) the tie breaking, and (d) the algorithm's output.}
  \label{fig:exalg}
\end{figure*}
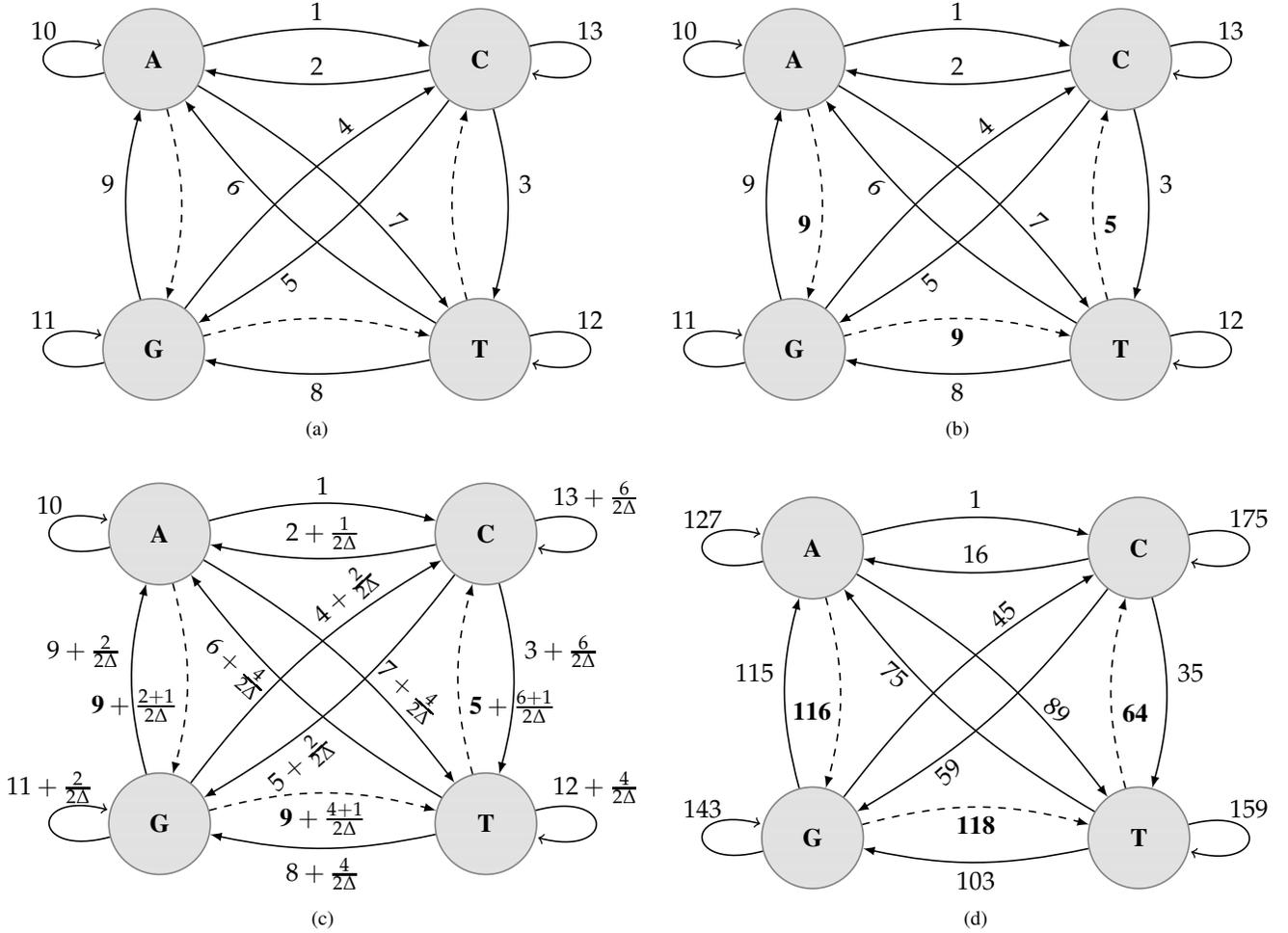

\begin{example}
  \label{ex:alg}
  We demonstrate Algorithm~\ref{alg:A} in action. Assume
  $\Sigma=\set{A,C,G,T}$, hence, $q=4$. Additionally, fix the window
  size as $\ell = 2$. Algorithm~\ref{alg:A} makes use of a
  predetermined Hamiltonian cycle, which we arbitrarily fix to be the
  one that is described by $AGTC$, namely, $\alpha=A\to G,G\to T,T\to
  C,C\to A$. Also, the algorithm requires an Eulerian cycle (whose
  prefix is $\alpha$), which we arbitrarily fix as the cycle described
  by the string $AGTCAACCTTATGGCG$ (namely, $\alpha,\beta=A\to G, G\to
  T, \dots$).

  Assume the user supplies the following input permutation:
  \[ \pi = \begin{pmatrix}
    AA & AC & AT & CA & CC & CG & CT & GA & GC & GG & TA & TG & TT \\
    9 & 0  & 8  & 1  & 12 & 4  & 2  & 8  & 3  & 10 & 5  & 7  & 11
  \end{pmatrix}. \]
  The main steps of the algorithms are:
  \begin{enumerate}
  \item
    Weights taken directly from $\pi$ are assigned to edges. This is
    shown in Figure~\ref{fig:case:1}. The dashed edges are the
    Hamiltonian path, and at this point, their weight has not been
    determined yet.
  \item
    Next, the algorithm determines the weights on the Hamiltonian path
    so that the graph becomes balanced. The result is shown in
    Figure~\ref{fig:case:2}.  We notice that two ties form: the weight
    of $A\to G$ equals that of $G\to T$, and the weight of $T\to C$
    equals that of $C\to G$.
  \item
    The algorithm then proceeds to break all ties. First, ties within
    $\alpha$ are broken. We assume here the algorithm does no
    optimization when finding $s$ and $s'$ in $\beta$, and simply
    takes the first occurrence satisfying the requirements. Thus, for
    $A\to G$ the algorithm uses $G\to G\to C\to G\to A$, for $G\to T$
    it uses $T\to T\to A\to T\to G$, and for $T\to C$ it uses $C\to
    C\to T$. Only then ties between $\alpha$ and $\beta$ are
    broken. The result is shown in Figure~\ref{fig:case:3}.
  \item
    The weights are made integers by multiplying by
    $2\Delta=14$. Finally, the weights are shifted so that the minimal
    weight is $1$, in this, reducing all weights by $13$. The end
    result, and algorithm output, is shown in Figure~\ref{fig:case:4}.
  \end{enumerate}
  
  We can see that at the end of the process we are left with weights
  that induce a permutation on the edges while preserving the order
  induced by the input permutation given by the user.
\end{example}

The size, and asymptotic rate of the code described in
Theorem~\ref{th:inject} is presented in the following corollary.

\begin{corollary}
  For all $q\geq 3$ and $\ell\geq 2$, there exists a
  $[q^\ell,q^\ell-q^{\ell-1}+1]$-systematic code
  $\cC_{q,\ell}\subseteq \Phi_{q,\ell}$. Additionally, the number of
  feasible permutations is lower bounded by
  \[F_{q,\ell}\geq \abs{\cC_{q,\ell}} = (q^{\ell}-q^{\ell-1}+1)!,\]
  and asymptotically the rate of feasible permutations satisfies
  \begin{align}
    \lim_{\ell\to\infty} R_{q,\ell} & \geq \lim_{\ell\to\infty} \frac{\log_2\abs{\cC_{q,\ell}}}{\log_2 (q^\ell!)} = 1-\frac{1}{q} \label{eq:sysell}\\
      \lim_{q\to\infty} R_{q,\ell} & \geq \lim_{q\to\infty} \frac{\log_2\abs{\cC_{q,\ell}}}{\log_2 (q^\ell!)} = 1. \label{eq:sysq}
  \end{align}
\end{corollary}
\begin{IEEEproof}
  The existence of $\cC_{q,\ell}$ with these parameters is immediate
  from Theorem~\ref{th:inject}, being the image of the injective
  mapping described there. For the asymptotic form, we recall
  Stirling's approximation, $\ln (n!) = n\ln (n)+ O(n)$ (e.g.,
  see~\cite[p.~452]{GraKnuPat94}).  With that we have
  \[
  R_{q,\ell} \geq \frac{\log_2 \abs{\cC_{q,\ell}}}{\log_2(q^\ell !)}
  = \frac{\log_2((q^{\ell}-q^{\ell-1}+1)!)}{\log_2(q^\ell !)}
  = \frac{(q^{\ell}-q^{\ell-1}+1)\log_2(q^{\ell}-q^{\ell-1}+1) + O(q^{\ell})}{q^\ell\log_2(q^\ell)+O(q^\ell)},\]
  and the claims follow.
\end{IEEEproof}

At this point we pause to compare our results with the best known,
described in~\cite{RavSchYaa19}. For $q\geq 3$ and $\ell\geq 2$, a
non-systematic code $\cC'_{q,\ell}\subseteq \Phi_{q,\ell}$ was
constructed in~\cite{RavSchYaa19}, for which
\[ \abs{\cC'_{q,\ell}}=30240\cdot\prod_{j=4}^q \parenv*{j!\cdot\binom{j^2-j+1}{j}}\cdot \prod_{i=3}^\ell (q!)^{q^{i-1}-2q^{i-2}+q^{i-3}}.\]
Apart for the case of $q=3$ and $\ell=2$ in which, in which our new
code is smaller,
\[\abs{\cC_{3,2}}=5040<30240=\abs{\cC'_{3,2}},\]
for all other cases, our new code is larger,
\[\abs{\cC_{q,\ell}} > \abs{\cC'_{q,\ell}}.\]
It should be emphasized that no explicit construction was presented
in~\cite{RavSchYaa19} for $q=3$ and $\ell=2$. Since this case was the
basis for a recursive construction, the size $30240$ was obtained
via an exhaustive computer search for all feasible permutations.
Asymptotically, as noted in~\cite{RavSchYaa19},
\[ \lim_{q\to\infty} \frac{\log_2\abs{\cC'_{q,\ell}}}{\log_2 (q^\ell!)} = \frac{1}{\ell},\]
which is out-performed by our results in~\eqref{eq:sysq}. More
importantly, in practical settings $q$ is fixed while
$\ell\to\infty$. In this asymptotic regime, the code
of~\cite{RavSchYaa19} gives
\[ \lim_{\ell\to\infty} \frac{\log_2\abs{\cC'_{q,\ell}}}{\log_2 (q^\ell!)} = 0,\]
which is inferior to our results in~\eqref{eq:sysell} that show a
non-vanishing rate.

\subsection{Upper Bound}

Having found a construction of systematic codes for feasible
permutation, it is natural to ask how large such systematic codes can
be. We provide an answer in the following theorem.

\begin{theorem}
  Let $q\geq 3$ and $k\geq 2$ be integers, and assume there exists a
  $[q^\ell,k]$-systematic code $\cC\subseteq\Phi_{q,\ell}$. Then,
  \[ k\leq q^{\ell}-q^{\ell-1} +1.\]
\end{theorem} 

\begin{IEEEproof}
  Assume to the contrary $k>q^{\ell}-q^{\ell-1} +1$, and let
  $G_{q,\ell-1}=(V,E)$ be the De Bruijn graph and $E'\subseteq E$ be
  an information set of size $\abs{E'}=k$. If we look at
  $G'=(V,E\setminus E')$, and forget the edge directions, then we have
  a graph with $q^{\ell-1}$ vertices, and strictly less than
  $q^{\ell-1}-1$ edges. This implies there exists an isolated vertex
  in $G'$, say $v\in V$. It then follows that
  \[ \Ein(v)\cup\Eout(v) \subseteq E',\]
  namely, all of the edges entering or leaving $v$ are part of the
  information set.

  By the definition of systematic codes,
  \[ \set*{\pi|_{E'} ; \pi\in\cC} = S_{E'}.\]
  In particular, there exists $\pi\in\cC$ such that $\pi(e)\leq
  \pi(e')$ for all $e\in\Ein(v)$ and $e'\in\Eout(v)$. However, $\pi$
  is clearly not feasible since we cannot balance $v$ when the weights
  of all incoming edges are smaller than the weights of all outgoing
  edges. Thus, we have reached a contradiction.
\end{IEEEproof}

\begin{corollary}
  \label{cor:opt}
  The systematic codes from Theorem~\ref{th:inject} are optimal.
\end{corollary}

\subsection{String Length}

An important figure of merit is the length of the string that the
encoder generates. We would like this string to be as short as
possible, to facilitate its synthesis. Thus, in this section we would
like to derive an upper bound on the maximal length of the string that
is generated by our algorithm. For any $q\geq3$ and $\ell\geq2$ we
will show that this length is polynomial in the input length. A
comparison with~\cite{RavSchYaa19} will show significant improvement.

Recall that Algorithm~\ref{alg:A} produces a profile vector, being the
weights of the De Bruijn graph $G_{q,\ell-1}$. The length of
associated string is simply the total weight of all edges. We now
prove an upper bound on this total weight. First, the following lemma
bounds the weight of an edge on the Hamiltonian path that is used by
the algorithm, before tie breaking. Since this weight might be
negative, we upper bound its absolute value.

\begin{lemma}
  \label{lem:HmEd}
  Let $G_{q,\ell-1}=(V,E)$ be the weighted De Bruijn graph after the
  balancing done in Algorithm~\ref{alg:A}, but before breaking
  ties. Let $E'_H$ be the set of edges in $G_{q,\ell-1}$ defined in
  Algorithm~\ref{alg:A}, and forming a Hamiltonian path. Then for each
  $e\in E'_H$,
  \[ \abs*{\wt(e)} \leq \wt(E\setminus E'_H) = \sum_{i=1}^{q^\ell-q^{\ell-1}+1}i=\binom{q^{\ell}-q^{\ell-1}+2}{2}\leq \frac{1}{2}q^{2\ell}.\]
\end{lemma}
\begin{IEEEproof}
  We use the notation of Algorithm~\ref{alg:A}. Assume
  $E'_H\eqdef\set{e_0,\dots,e_{q^{\ell-1}-2}}$ and
  $e_0,e_1,\dots,e_{q^{\ell-1}-2}$ is a Hamiltonian path in
  $G_{q,\ell-1}$, where $e_i$ is the edge $v_i\to v_{i+1}$.  For all
  $i\in[q^{\ell-1}-1]$, we define $U_i\eqdef\set{v_0,v_1,\dots,v_i}$,
  and we observe that
  \[
  E'_H\cap \Ein(U_i)=\emptyset, \qquad\text{and}\qquad
  E'_H\cap \Eout(U_i)=\set{e_i}.
  \]
  By Lemma~\ref{lem:balset} we get that
  \[ \wt(e_i) = \wt(\Ein(U_i))-\wt(\Eout(U_i)\setminus\set{e_i}).\]
  The claim is now immediate, since both $\Ein(U_i)\subseteq
  E\setminus E'_H$ and $\Eout(U_i)\setminus\set{e_i}\subseteq
  E\setminus E'_H$.
\end{IEEEproof}

\begin{theorem}
  \label{th:len}
  Let $G_{q,\ell-1}=(V,E)$ be the weighted De Bruijn graph that is the output
  of Algorithm~\ref{alg:A}. Then
  \[ \wt(E) \leq q^{5\ell }.\]
\end{theorem}

\begin{IEEEproof}
  We again use the notation of Algorithm~\ref{alg:A}. Recall that all
  the edges in $E\setminus E'_H$ are initially given distinct weights
  from $\set{1,\dots,q^{\ell}-q^{\ell-1}+1}$, whose sum is upper
  bounded by $q^{2\ell}/2$, as in Lemma~\ref{lem:HmEd}. Again, by
  Lemma~\ref{lem:HmEd}, the weight of any $e\in E'_H$ satisfies
  $\abs{\wt(e)}\leq q^{2\ell}/2$, before breaking ties. After the
  algorithm breaks all ties, the weight of each edge is increased by
  no more than $1$. Then all weights are multiplied by
  $2\Delta=2(\binom{q^{\ell-1}}{2}+1)\leq q^{2\ell-2}/2$. Finally, the
  normalization process may decrease or increase the weight of all
  edges. If an increase occurs, that it is only because some edge in
  $E'_H$ has negative weight. Thus, a weight of no more than
  $q^{2\ell}/2$ is added to all edges. It follows that the total
  weight of the output weighted graph satisfies,
  \[ \wt(E) \leq 2\Delta\parenv*{\frac{q^{2\ell}}{2}+(q^{\ell-1}-1)\cdot\frac{q^{2\ell}}{2}+q^{\ell}\cdot 1}+q^\ell\cdot\frac{q^{2\ell}}{2}\leq q^{5\ell}, \]
  as claimed.
\end{IEEEproof}

We first comment that the bound of Theorem~\ref{th:len} may be
improved by a constant factor by having a more careful analysis in
Lemma~\ref{lem:HmEd}, taking into account the maximal cut size in
$G_{q,\ell-1}$, as well as finer inequalities in
Theorem~\ref{th:len}. However, recognizing the fact that we are
interested in the asymptotic regime where $q$ is constant and
$\ell\to\infty$, the resulting upper bound is still $O(q^{5\ell})$.

Putting our results in context, if we denote the length of the input
to Algorithm~\ref{alg:A} by $N\eqdef q^{\ell}-q^{\ell-1}+1$, then the
upper bound of Theorem~\ref{th:len} is $O(N^5)$. Thus,
Algorithm~\ref{alg:A} guarantees an output string length that is
polynomial in the input length. Additionally, the absolute minimum
string length is lower bounded by the case of assigning the weights
$\set{1,2,\dots,q^{\ell}}$ to the edges, giving a lower bound of
\[\sum_{i=1}^{q^\ell} i = \binom{q^{\ell}+1}{2} = \Omega(q^{2\ell})=\Omega(N^2).\]

Finally, we would like to compare our upper bound on the length of the
output string from Algorithm~\ref{alg:A}, to the upper bound on the
length of the output string from the encoding algorithms
in~\cite{RavSchYaa19}. For general $q\geq 3$ and $\ell\geq 2$, it was
shown in~\cite{RavSchYaa19} that the upper bound is
\[ 16\cdot \frac{2^{q-4}\cdot q!\cdot (q+1)!}{144\cdot q^2}\cdot 3^{\ell-2}\cdot q^{\ell(q^2+1)}=O(3^{\ell}q^{\ell(q^2+1)}),\]
in the asymptotic regime of constant $q$ and $\ell\to\infty$. This
bound is worse than that of Theorem~\ref{th:len}.

\begin{remark}
  \label{rem:l2}
  When $\ell=2$, the phase of tie-breaking in $\alpha$ in
  Algorithm~\ref{alg:A} takes on a very simple form. This is because
  for every edge $e_i$, $1\leq i\leq q^{\ell-1}-1$, the reverse edge
  exists in the graph, and is not part of $\alpha$. Thus, all the
  cycles used in this phase may be chosen to be edge disjoint, and
  then $\Delta$ may be reduced to $\Delta=q^{\ell-1}$. In that case,
  the bound on the output-string length of Theorem~\ref{th:len} becomes
  $\wt(E)\leq \frac{3}{2}q^6+2q^3$.
\end{remark}

For the specific case of $\ell=2$, \cite{RavSchYaa19} showed an upper
bound of $q^2\cdot 2^{q-3}\cdot
\frac{q!}{6}\cdot\frac{(q+1)!}{24}\cdot 16$, which for $q=3$ is better
than the bound in Remark~\ref{rem:l2}, but is otherwise worse.

\section{Non-systematic Codes}
\label{sec:nonsys}

In the previous section we studied systematic rank-modulation codes,
and we attained the maximum possible rate
(Corollary~\ref{cor:opt}). In this section we drop this constraint,
and show that there are significantly larger codes that are
non-systematic. Since the number of feasible permutation is still
unknown, comparing our results with the optimum is impossible, and
instead we compare against the systematic codes of the previous
section.

Our first observation is a trivial increase in the code size, by using
the self-loop edges in the De Bruijn graph.

\begin{lemma}
  \label{lem:selfloop}
  For all $q\geq 3$ and $\ell\geq 2$, there exists a $(q,M)$-code
  $\cC\subseteq\Phi_{q,\ell}$, with
  $M=(q^\ell-q^{\ell-1}+1-q)!\frac{q^\ell!}{(q^\ell-q)!}$.
\end{lemma}

\begin{IEEEproof}
  Remove the $q$ self-loop edges from the De Bruijn graph, and run
  Algorithm~\ref{alg:A}. We note that removing these edges does not
  affect the algorithm in any way. Then, set the weight of the
  self-loop edges arbitrarily. The weighted graph will remain
  balanced. The number of permutations obtained this way is the
  claimed value of $M$.
\end{IEEEproof}

Next, we explore new sufficient conditions and necessary conditions
for the existence of feasible permutations. We begin with a simple
extension of a necessary condition presented
in~\cite{RavSchYaa19}. Since~\cite{RavSchYaa19} used the vertices of
the De Bruijn graph, whereas here we balance edge weights, we require
the following new definition.

\begin{definition}
  Assume $q\geq 3$, $\ell\geq 2$, and let $G_{q,\ell-1}=(V,E)$ be a
  weighted balanced De Bruijn graph. Let $\emptyset\subset U\subset
  V$ be a non-empty proper subset of $V$, and assume
  \[\Ein(U)=\set{e_0,\dots,e_{k-1}}, \qquad \Eout(U)=\set{e'_0,\dots,e'_{k-1}},\]
  are indexed such that
  \[ \wt(e_0)<\dots<\wt(e_{k-1}), \qquad \wt(e'_0)<\dots<\wt(e'_{k-1}).\]
  We say $U$ exhibits a \emph{Dyck configuation} if either
  \[ \wt(e_i)<\wt(e'_i) \quad \text{for all $i\in [k]$,}\]
  or
  \[ \wt(e'_i)<\wt(e_i) \quad \text{for all $i\in [k]$.}\]
  Additionally, we say a permutation $\pi\in S_{E'}$,
  $\Ein(U)\cup\Eout(U)\subseteq E'\subseteq E$, exhibits a Dyck
  configuration at $U$, if setting $\wt(e)=\pi(e)$ for all $e\in E'$,
  creates a Dyck configuration at $U$.
\end{definition}

Assume the edges in $\Ein(U)\cup\Eout(U)=\set{e_0,\dots,e_{2k-1}}$ are
indexed such that $e_0<e_1<\dots<e_{2k-1}$. We can construct the
following binary word $b_0, b_1, \dots, b_{2k-1}$, where $b_i$ is $0$
if $e_i\in\Ein(U)$, and is $1$ otherwise. This word is a Dyck
word\footnote{A Dyck word is a binary sequence, exactly half of its
  bits are $0$'s, such that, in each of its prefixes, the number of
  $0$'s is at least the number of $1$'s.}  if and only if $U$ exhibits
a Dyck configuration. We now present a necessary condition for a
permutation to be feasible.

\begin{lemma}
  \label{lem:comb}
  Let $q\geq 3$ and $\ell\geq 2$. If $\pi\in \Phi_{q,\ell}$ is a feasible
  permutation, then for all $\emptyset\subset U\subset
  \Sigma^{\ell-1}$, $\pi$ does not exhibit a Dyck configuration at $U$.
\end{lemma}

\begin{IEEEproof}
  Let $G_{q,\ell-1}=(V,E)$ be the De Bruijn graph, and set
  $\wt(e)=\pi(e)$ for all $e\in E$. Assume to the contrary that $\pi$
  is feasible but there exists $\emptyset\subset U\subset
  V=\Sigma^{\ell-1}$ that exhibits a Dyck configuration. It follows
  that either $\wt(\Ein(U))<\wt(\Eout(U))$, or
  $\wt(\Ein(U))>\wt(\Eout(U))$. Since $\pi$ is feasible, there exists
  a feasible $x\in\N^{\Sigma^\ell}$ such that $x\vDash\pi$. It then
  follows that, either
  \[ \sum_{e\in\Ein(U)} x(e) < \sum_{e\in\Eout(U)} x(e), \quad\text{or}\quad
  \sum_{e\in\Ein(U)} x(e) > \sum_{e\in\Eout(U)} x(e). \]
  However, the fact that $x$ is feasible implies, by Lemma~\ref{lem:balset},
  that
  \[ \sum_{e\in\Ein(U)} x(e) = \sum_{e\in\Eout(U)} x(e), \]
  a contradiction.
\end{IEEEproof}

The necessary condition for a permutation to be feasible, which was
presented in Lemma~\ref{lem:comb}, is unfortunately not a sufficient
condition, as the following example shows.

\begin{example}
  Take $q=4$ with $\Sigma=\set{A,C,G,T}$, and $\ell=2$. Consider the
  following permutation:
  \[
  \pi=\begin{pmatrix}
  AA & AC & AG & AT & CA & CC & CG & CT & GA & GC & GG & GT & TA & TC & TG & TT\\
  12 & 0  & 1  & 5  & 4  & 13 & 11 & 7  & 3  & 10  & 14 & 6 & 2  & 8  & 9  & 15
  \end{pmatrix}.
  \]
  By inspection, one can verify that no $\emptyset\subset U\subset
  \Sigma$ exhibits a Dyck configuration. However, by computer we find
  that this permutation is infeasible (see the linear-programming
  method for deciding feasibility in~\cite[Section IV]{RavSchYaa19}).
\end{example}

Not all is lost though. In the next theorem we show that, compared
with the systematic code of Theorem~\ref{th:inject}, the user may set
another edge, provided that a Dyck configuration does not appear. To
prove this claim we require a little preparation.

\begin{definition}
  Let $G=(V,E)$ be a finite directed weighted graph. A vertex $v\in V$ is said
  to be in an \emph{over} (respectively, \emph{under}) state, if
  $\wt(\Ein(v))<\wt(\Eout(v))$ (respectively,
  $\wt(\Ein(v))>\wt(\Eout(v))$). Otherwise, $v$ is said to be
  \emph{balanced}.
\end{definition}

\begin{definition}
  Let $G=(V,E)$ be a finite directed weighted graph. For any $e\in\E$,
  define
  \[ E_{\geq e} \eqdef \set*{ e'\in E ; \wt(e')\geq \wt(e) }.\]
  We say $e^*\in E$ is a \emph{step-up edge for $v\in V$} if
  $\abs*{E_{\geq e^*}\cap \Ein(v)}<\abs*{E_{\geq e^*}\cap
    \Eout(v)}$. We say $e^*\in E$ is a \emph{step-down edge for $v\in
    V$} if $\abs*{E_{\geq e^*}\cap \Ein(v)}>\abs*{E_{\geq e^*}\cap
    \Eout(v)}$. Finally, we say $e^*\in E$ is a \emph{stable edge for
    $v\in V$} if $\abs*{E_{\geq e^*}\cap \Ein(v)}=\abs*{E_{\geq
      e^*}\cap \Eout(v)}$.
\end{definition}

Unlike Lemma~\ref{lem:keepbal}, we introduce an operation that may
change the balanced state of vertices.

\begin{lemma}
  \label{lem:calibrate}
  Let $G=(V,E)$ be a finite directed weighted graph. Assume that $v\in
  V$ is in an over state (resp., in an under state), and that $e^*\in
  E$ is a step-down edge (resp., step-up edge) for $v$. Construct
  $G'=(V,E)$, and set its edge weights as follows:
  \[ \wt_{G'}(e)=\begin{cases}
  \wt_G(e) & e\notin E_{\geq e^*}, \\
  \wt_G(e)+c & e\in E_{\geq e^*},
  \end{cases}
  \]
  for all $e\in E$, and where
  \[c=\frac{\abs*{\wt_G(\Ein(v))-\wt_G(\Eout(v))}}{\abs*{\abs*{E_{\geq e^*}\cap \Ein(v)}-\abs*{E_{\geq e^*}\cap \Eout(v)}}}.\]
  Then the relative order of edges (by weight) in $G$ and $G'$ are the
  same, and $v$ is balanced in $G'$.
\end{lemma}

\begin{IEEEproof}
  The fact that the relative order of edges does not change between
  $G$ and $G'$, is trivial. Assume $v$ is in an under state, i.e.,
  $\wt_G(\Ein(v))>\wt_G(\Eout(v))$. Since $e^*$ is a step-up edge for
  $v$, by definition we have $\abs*{E_{\geq e^*}\cap
    \Ein(v)}<\abs*{E_{\geq e^*}\cap \Eout(v)}$. We now note that
  increasing the weights of the edges in $E_{\geq e^*}$ by $1$, adds
  $\abs{E_{\geq e^*}\cap\Ein(v)}$ weight to the incoming edges of $v$, and
  $\abs{E_{\geq e^*}\cap\Eout(v)}$ weight to the outgoing edges of $v$. Thus,
  \[
  \wt_{G'}(\Ein(v))-\wt_{G'}(\Eout(v)) = \wt_{G}(\Ein(v))+
  c\abs*{E_{\geq e^*}\cap\Ein(v)}-\wt_G(\Eout(v))-c\abs*{E_{\geq e^*}\cap\Eout(v)}=0.
  \]
  A symmetric argument proves the case when $v$ is in an over state.
\end{IEEEproof}

We are now in a position to show how another edge may be set (compared
with systematic codes), provided a Dyck configuration is avoided.

\begin{theorem}
  \label{th:firstnode}
  Assume the same setting as in Theorem~\ref{th:inject}. Then every
  permutation $\pi$ on $E\setminus E'_H\cup\set{e_0}$, that does not
  exhibit a Dyck configuration at $\set{v_1}$, can be extended to a
  feasible permutation $\pi'\in\Phi_{q,\ell}$, namely,
  $\pi'|_{E\setminus E'_H\cup\set{e_0}}=\pi$.
\end{theorem} 

\begin{IEEEproof}
  Our goal is to show that we can find weights $x(e)$ for each $e\in
  E$, such that the graph is balanced, and the relative order (by
  weight) of the edges in $E\setminus E'_H\cup\set{e_0}$ is
  preserved. We start by setting $x(e)=\pi(e)+1$ for each $e\in
  E\setminus E'_H\cup\set{e_0}$. We then note $v_0$ is the only vertex
  all of whose incident edge weight have already been set.

  If $v_0$ is not balanced, then it is either in an over state or an
  under state. Let us assume that $v_0$ is in an under state. The
  proof for the over state is symmetric. Arrange the edges of
  $\Ein(v_0)\cup\Eout(v_0)$ in ascending weight order,
  $e'_0,\dots,e'_{2k-1}$, where we note that by definition, self loops
  are not included in this union. Create the binary word
  $b_0,\dots,b_{2k-1}$, $b_i\in\set{0,1}$, where $b_i=0$ if and only
  if $e'_i\in\Ein(v_0)$. In this word exactly half of the bits are
  $0$'s. Since there is no Dyck configuration at $\set{v_0}$, there
  exists a proper prefix $b_0,\dots,b_{t-1}$ that contains strictly
  more $0$'s than $1$'s, and therefore, $b_t,\dots,b_{2k-1}$ contains
  strictly more $1$'s than $0$'s. Thus, $e'_t$ is a step-up edge
  for $v_0$.

  Using Lemma~\ref{lem:calibrate}, we can adjust the weights of edges
  (that have already been assigned) and ensure that $v_0$ is balanced,
  while keeping the relative order of edges by weight. We also point
  out that the resulting weights must all be distinct. If needed, we
  multiply all edge weights by the same constant to obtain integer
  weights. We now continue by running Algorithm~\ref{alg:A}, starting
  from the balancing part.  The resulting weights induce a permutation
  $\pi'\in\Phi_{q,\ell}$, as desired.
\end{IEEEproof}

\begin{corollary}
  \label{cor:nonsyssize}
  For all $q\geq 3$ and $\ell\geq 2$, there exists a code
  $\cC'_{q,\ell}\subseteq \Phi_{q,\ell}$ with
  \[F_{q,\ell}\geq \abs{\cC'_{q,\ell}} = (q^{\ell}-q^{\ell-1}+2-q)!\cdot\frac{q-1}{q+1}\cdot\frac{q^\ell!}{(q^\ell-q)!}.\]
\end{corollary}

\begin{IEEEproof}
  We choose $v_0$ in the setting of Theorem~\ref{th:firstnode} to be a
  vertex with no self loops. It follows that
  $\abs{\Ein(v_0)}=\abs{\Eout(v_0)}=q$. We first look locally at
  $v_0$. We can arrange the $q$ incoming edges among themselves in
  $q!$ ways, and similarly for the $q$ outgoing edges. Next, we count
  the number of ways these two orderings may be merged so as not to
  exhibit a Dyck configuration. A Dyck configuration is equivalent to
  a Dyck word, and the number of those is known to be the Catalan
  number $C_q\eqdef\frac{1}{q+1}\binom{2q}{q}$ (e.g.,
  see~\cite[p.~358]{GraKnuPat94}). There are also two ways to choose
  whether the first edge is from $\Ein(v_0)$ or $\Eout(v_0)$. We
  obtain that the total ways of ordering $\Ein(v_0)\cup\Eout(v_0)$ is
  given by
  \[ (q!)^2 \parenv*{\binom{2q}{q}-2C_q}=(2q)!\cdot\frac{q-1}{q+1}.\]
  We then extend this to a permutation of $E\setminus E'_H\cup\set{e_0}$,
  for a total number of permutations equalling
  \[ (q^\ell-q^{\ell-1}+2)!\cdot\frac{q-1}{q+1}.\]
  By Theorem~\ref{th:firstnode} these may be injectively extended to
  feasible permutations of $E$. As a final step, we may employ the
  same strategy as Lemma~\ref{lem:selfloop}: exclude the self loops
  from the entire process, and set their value only at the end.  This
  results in the claimed number of permutations in $\cC'_{q,\ell}$.
\end{IEEEproof}

We can further improve Theorem~\ref{th:firstnode}, by considering
permutations on \emph{all} the edges of the De Bruijn graph. A sufficient
condition is described in the following theorem.

\begin{theorem}
  \label{th:allnodes}
  Assume $q\geq 3$, $\ell\geq 2$, and let $G_{q,\ell-1}=(V,E)$ be the
  De Bruijn graph. Let $\pi\in S_E$, and assign $\wt(e)=\pi(e)+1$, for
  all $e\in E$. If we can index the vertices
  $V=\set{v_0,\dots,v_{q^{\ell-1}-1}}$ such that for each
  $i\in[q^{\ell-1}-1]$:
  \begin{enumerate}
  \item
    there is no Dyck configuration at $\set{v_i}$, and
  \item
    $v_i$ has a step-up edge $e\in E$, and a step-down edge $e'\in E$,
    such that $e$ and $e'$ are stable edges for $v_j$, for all $j\in[i]$,
  \end{enumerate}
  then $\pi$ is feasible, i.e., $\pi\in\Phi_{q,\ell}$.
\end{theorem} 

\begin{IEEEproof}
  When the conditions in the lemma are satisfied, we can balance each
  $v_0,\dots,v_{q^{\ell-1}-2}$, one by one, in this order, using
  Lemma~\ref{lem:calibrate}. Note that while we balance $v_i$, we do
  not harm the balance of $v_0,\dots,v_{i-1}$. Also note that vertex
  $v_{q^{\ell-1}-1}$ is automatically balanced once all the previous
  ones are. The result is a balanced graph with rational
  weights. Multiplying all the weights by an appropriate constant we
  achieve a balanced graph with distinct integer positive weights,
  that realize the permutation $\pi$.
\end{IEEEproof}

Alas, the sufficient condition for a permutation to be feasible, which was
presented in Theorem~\ref{th:allnodes}, is not necessary, as the
following example shows.

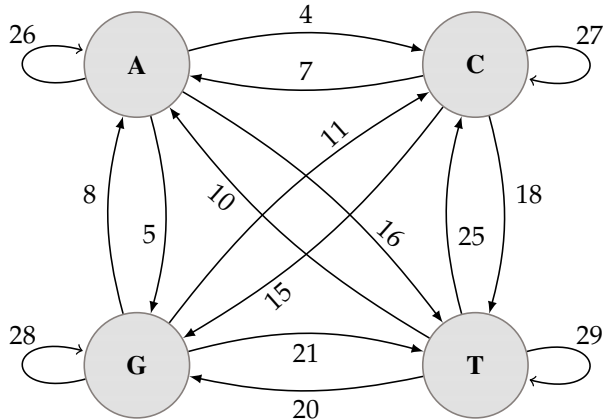
\begin{figure}[th]
  \begin{center}
    \definecolor {processblue}{cmyk}{0.3,0.3,0.3,0.3}
    \begin {tikzpicture}[-latex ,auto ,node distance =4 cm and 4.5cm ,on grid ,
        semithick ,
        state/.style = { circle ,top color =processblue!20 , bottom color = processblue!20 ,
          draw,processblue , text=black , minimum width =1.4 cm}]
      \node[state] (G){$\textbf{G}$};
      \node[state] (A) [above =of G] {$\textbf{A}$};
      \node[state] (T) [right =of G] {$\textbf{T}$};
      \node[state] (C) [above =of T] {$\textbf{C}$};
      \path (A) edge [loop left] node[above = 0.15] {$26$} (A);
      \path (G) edge [bend left =15] node[above left= 0.04 cm]{$8$} (A);
      \path (A) edge [bend right = -15] node[below left =0.001 cm] {$5$} (G);
      \path (A) edge [bend left=11, align=left, sloped, inner sep=1pt] node[right = 1,above ] {$16$} (T);
      \path (T) edge [bend left=11, align=left, sloped, inner sep=1pt] node[left = 1,below]{$10$} (A);
      \path (G) edge[bend left =15] node[below] {$21$} (T);
      \path (T) edge [bend left =12] node[below ] {$20$} (G);
      \path (T) edge [loop right] node[above = 0.15] {$29$} (T);
      \path (G) edge [loop left ] node[above = 0.15] {$28$} (G);
      \path (C) edge [loop right] node[above = 0.15] {$27$} (C);
      \path (C) edge [bend right = -15] node[above right =0.04 cm] {$18$} (T);
      \path (A) edge [bend left =15] node[above] {$4$} (C);
      \path (C) edge [bend left =12] node[above] {$7$} (A);
      \path (T) edge [bend left =15] node[below right= 0.04 cm] {$25$} (C);
      \path (C) edge [bend left=11, align=left, sloped, inner sep=1pt] node[left= 1,below ] {$15$} (G);
      \path (G) edge[bend left=11, align=left, sloped, inner sep=1pt] node[right= 1,above ] {$11$} (C);
    \end{tikzpicture}
  \end{center}
  \caption{The balanced graph for the permutation from Example~\ref{ex:notnec}.}
  \label{fig:notnec}
\end{figure}

\begin{example}
  \label{ex:notnec}
  Take $q=4$ with $\Sigma=\set{A,C,G,T}$, and $\ell=2$. Consider the
  following permutation:
  \[
  \pi=\begin{pmatrix}
  AA & AC & AG & AT & CA & CC & CG & CT & GA & GC & GG & GT & TA & TC & TG & TT\\
  12 & 0  & 1  & 7  & 2  & 13 & 6  & 8  & 3  & 5  & 14 & 10 & 4  & 11 & 9  & 15
  \end{pmatrix},
  \]
  By inspection one can verify that the requirements of
  Theorem~\ref{th:allnodes} are not satisfied, yet the permutation is
  feasible, as Figure~\ref{fig:notnec} shows.
\end{example}

Table~\ref{tab:comparison} shows a comparison between the size of the
codes resulting from the different methods in this paper and
in~\cite{RavSchYaa19}.  We first note that the last row, the total
number of feasible permutations, was obtained using an exhaustive
computer search, and hence the limitation to $\ell=2$ and $q=3,4$. We
also observe that the entry for $q=3$, $\ell=2$,
from~\cite{RavSchYaa19} was obtained in the same way, i.e., an
exhaustive computer search, whose results bootstrapped a recursive
construction in~\cite{RavSchYaa19}.

\begin{table}[th]
  \caption{The rate of rank-modulation codes for DNA storage with
  shotgun sequencing (in parentheses, the code rates)}
  \label{tab:comparison}
  \[
  \begin{array}{r|c|c}
    \text{Source} & \ell=2, q=3 & \ell=2, q=4  \\ \hline
    \text{\cite{RavSchYaa19}} & 30240 \ (\approx 0.806) & 518918400 \ (\approx 0.654) \\
    \text{Theorem~\ref{th:inject} (Algorithm~\ref{alg:A})} & 5040 \ (\approx 0.666)& 6227020800 \ (\approx 0.735)\\
    \text{Theorem~\ref{th:firstnode}} & 30240 \ (\approx 0.806)& 95103590400 \ (\approx 0.824)\\
    \text{Theorem~\ref{th:allnodes}} & 30240 \ (\approx 0.806)& 1296453150720 \ (\approx 0.909)\\
    \text{Total feasible permutations} & 30240 \ (\approx 0.806)& 1540034496000 \ (\approx 0.915)\\
  \end{array}
  \]
\end{table}

\section{Conclusion}
\label{sec:conc}

In this paper we studied rank-modulation codes for DNA storage when
used in conjunction with shotgun sequencing. We constructed systematic
codes for all parameters $q\geq 3$ and $\ell\geq 2$, which we proved
are optimal. These improve upon the results of~\cite{RavSchYaa19} by
obtaining an asymptotic rate of $1-\frac{1}{q}$ when $q$ is fixed and
$\ell\to\infty$, compared with an asymptotic rate of $0$
in~\cite{RavSchYaa19}. In the asymptotic regime of $\ell$ fixed and
$q\to\infty$ we obtain asymptotic rate of $1$ compared with
$\frac{1}{\ell}$ in~\cite{RavSchYaa19}. Finally, we also showed how
larger codes may be obtained by avoiding Dyck configurations.

We would like to further discuss additional aspects that may be
readily combined into the coding schemes we presented in this paper:

\paragraph{Weight Balancing}
When considering data storage in synthesized DNA molecules, it has
been argued that an overall GC-content\footnote{The GC-content of a
  DNA molecule is the percentage of bases that are either $G$ or $C$.}
of roughly $50\%$ contributes to the stability of the
molecule~\cite{YakProFra06}. We can adjust Algorithm~\ref{alg:A} to
accomplish this by removing the self loops from the set of information
edges in the De Bruijn graph $G_{q,\ell-1}$. After running the
algorithm and obtaining an encoded sequence, we may increase the
weights of the relevant self loops to reach the desired GC-content of
the encoded sequence.

\paragraph{Forbidden $\ell$-grams}
Research suggests that some $\ell$-grams are likely to cause
sequencing errors~\cite{Nak11}. We can ensure these $\ell$-grams never
appear as a substring of the encoded output sequence by removing their
corresponding edges from the De Bruijn graph $G_{q,\ell-1}$ to obtain
a graph $G'$. A careful reading of Theorem~\ref{th:inject} and
Algorithm~\ref{alg:A} reveals that the claims hold for $G'$ as well,
provided the following hold:
\begin{itemize}
\item
  $G'$ has an Eulerian cycle.
\item
  $G'$ has a Hamiltonian path.
\item
  For each edge $e$ on the Hamiltonian path, there is a directed cycle
  passing through $e$ and not through any of the other edges on the
  Hamiltonian path.
\end{itemize}
When these requirements hold, the edges not on the Hamiltonian path
form an information set, and trivial adjustments to
Algorithm~\ref{alg:A} make it work for $G'$ as well.

\paragraph{Error Correction}
As mentioned in the introduction, the mere use of the rank-modulation
scheme already protects against perturbations of the profile vector
that do not change the ranking. If we desire more error-protection
capabilities, then we may use any of the rank-modulation
error-correcting codes known in the literature. These may be trivially
combined with the systematic encoding of Section~\ref{sec:sys}. In the
notation of Theorem~\ref{th:inject}, if $C\subseteq S_{E\setminus
  E'_H}$ is a rank-modulation code, then each of its codewords may be
mapped to $\Phi_{q,\ell}$. The reverse process is easily accomplished
by projecting the receiving permutation onto $E\setminus E'_H$, and
then decoding using $C$. We can also use the larger non-systematic
codes from Section~\ref{sec:nonsys}. Assume $\Cout\subseteq
S_{q,\ell}$ is the (non-systematic) code from
Section~\ref{sec:nonsys}, and let $\Cin\subseteq S_{q,\ell}$ be a
rank-modulation error-correcting code. If $\Cin$ is a subgroup code
(e.g., the codes studied in~\cite{TamSch10}), then its cosets
partition $S_{q,\ell}$ into error-correcting codes (in the case
of~\cite{TamSch10}, due to the right-invariance of the
$\ell_\infty$-metric on permutations). Thus, one of these cosets
intersects $\Cout$ in a code that is both feasible, has the
error-correction capabilities of $\Cin$, and whose size is at least
$\abs{\Cout}\cdot\abs{\Cin}/\abs{S_{q,\ell}}$.

We would like to mention some open questions. Finding the exact number
of feasible permutations is first and foremost. The upper bound on the
asymptotic rate of feasible permutations is still $1$
(see~\cite{RavSchYaa19}), whereas the lower bound has been improved in
this paper to $1-\frac{1}{q}$, assuming $q$ is constant and
$\ell\to\infty$. This lower bound is obtained by considering
systematic codes, and it is the best possible. Unfortunately, even
though the non-systematic codes of Corollary~\ref{cor:nonsyssize} have
strictly larger size compared with systematic codes, they do not offer
any improvement asymptotically.

Another interesting open question concerns the length of the encoded
sequences. The trivial lower bound is $\Omega(q^{2\ell})$, whereas the
upper bound from the systematic codes of Section~\ref{sec:sys} is
$O(q^{5\ell})$. What are the worst-case bound and the average-case
bound is still unknown.

Yet another open problem is determining the minimum distance of
feasible permutations. Several metrics have been studied in connection
with rank-modulation codes, e.g., Kendall's $\tau$-metric, the
$\ell_\infty$-metric (also known as Chebyshev's metric), and Ulam's
metric, to name a few.  Intrinsically, the set of feasible permutations
may possess sufficient minimal distance to allow error
correction. What this distance is, or bounds on it, are as of yet,
unknown.

Finally, finding a concise sufficient and necessary condition for a
permutation to be feasible, remains an open problem. Finding such a
condition might pave the way to constructing encoders for feasible
permutations. We leave all of these open problems for future work.

\bibliographystyle{IEEEtranS}
\bibliography{allbib}

\end{document}